\documentclass {statsoc}
\usepackage[a4paper]{geometry}

\usepackage [utf8] {inputenc}

\usepackage{etoolbox}

\makeatletter
\patchcmd{\@makecaption}
  {\parbox}
  {\advance\@tempdima-\fontdimen2} 
  {}{}
\makeatother  

\usepackage {graphicx}
\usepackage {amsfonts}
\usepackage {amsmath}

\usepackage [english]{babel}
\usepackage [autostyle,english = american]{csquotes}
\MakeOuterQuote{"}

\usepackage {natbib}
\usepackage{hyperref}

\newtheorem{lemma}{LEMMA}
\newtheorem{theorem}{THEOREM}

\usepackage[dvipsnames]{xcolor}

\usepackage{color}

\title{FATSO: A family of operators for variable selection in linear models}
\date{}
\author{Nicolás E. Kuschinski}
\author{J. Andrés Christen}
\address{CIMAT, Guanajuato, Gto., Mexico}
\begin{document}
\maketitle{}

\begin{abstract}
In linear models it is common to have situations where several regression coefficients are zero. In these situations a common tool to perform regression is a variable selection operator. One of the most common such operators is the LASSO operator, which promotes point estimates which are zero. The LASSO operator and similar approaches, however, give little in terms of easily interpretable parameters to determine the degree of variable selectivity. In this paper we propose a new family of selection operators which builds on the geometry of LASSO but which yield an easily interpretable way to tune selectivity. These operators correspond to Bayesian prior densities and hence are suitable for Bayesian inference.  We present some examples using simulated and real data, with promising results.

\end{abstract}

\section{Introduction}

In standard linear models, it is not uncommon to have prior knowledge that several of the regression coefficients should be zero.  This happens, for example, when it is suspected that most of the factors considered in a large model are not relevant.  The identity of which factors \textit{are} relevant, however, is not known beforehand.  It is of interest to estimate model parameters and to identify which coefficients are nonzero. This is a classical problem with a classical solution, wherein regression is performed on the model and then the parameters are tested one at a time to determine whether they are significantly nonzero. Often this is followed by a second round of inference using only those parameters determined to be nonzero the first time through \citep[see][for example]{linearreg}.

This procedure works well for large sample sizes and small dimensions, but for high dimensions (large numbers of explanatory variables $X$) or small samples sizes, eg. the $n > p$ problem, it becomes impossible to perform linear regression using classical techniques since the response $Y$ is typically found exactly inside of the column space of $X$, and there are infinitely many exact solutions.

There are several solutions to this problem, but many of them center only around estimation and do not intend to identify relevant factors. Methods that do intend to separate relevant from irrelevant variables are known as \textit{variable selection} methods. There is a broad body of recent literature on the subject of variable selection in extremely high dimensional problems, such as those which are frequently encountered in gene selection and microarray data \citep{variablesel}. In this paper we will focus on linear regression problems, usually with a more manageable (if still large) number of dimensions.

In order to obtain good estimates in these situations, one common solution is to use regularizing operators. One popular such operator is the LASSO operator \citep{tibshirani}, which is designed to yield point estimates which are frequently exactly zero. Tuning the degree of selectivity of the LASSO operator, however, is not very fluid. The degree of selectivity is tied to shrinkage of the estimators and it is difficult to interpret.

While LASSO is a very popular operator for variable selection in linear models (and has been tried in non linear models also, see \citealp {nonlinlasso}), several other regularization methods exist, such as ridge regression \citep{ridge}, bridge regression \citep{bridge}, elastic net \citep{elasticnet}, etc.  While these operators have several important virtues, none of them address the issue of interpretability in variable selection.

In a Bayesian setting, a large number of selection operators have been suggested recently in the form of scale mixtures of normals.  For instance the horseshoe prior \citep {horseshoe}, Dirichlet-Laplace priors \citep {otherscalemix} and others \cite{mixgpriors}. These approaches also do not focus on interpretability issues.  Our selection operator is not scale a mixture of normals; we follow a different strategy.

In this paper, we look into the geometric mechanism by which LASSO promotes regression estimators to zero, and we study some of the consequences. Using this information we propose a new family of operators which use the same geometric mechanism as LASSO, but provide an extra parameter which permits fluid and intuitive tuning of the degree of selectivity separately from shrinkage.
We prove that this family of operators corresponds to a large family of Bayesian prior distributions, and we study the relationship between the geometry of the priors and the meaning of the parameters in a Bayesian context.

The paper is organized as follows.  In section \ref{lassosec} we establish our notation and investigate the geometric properties of the LASSO operator. In section \ref{introfatso} we introduce the FATSO operator, which is based on the geometrical properties discussed in section \ref{lassosec}. In section \ref{rho} we study the behavior of FATSO and see how it addresses the issue of parameter interpretability. Section \ref{comparisons} discusses the differences between FATSO and various other extensions of LASSO. In section \ref{resultsec} we look into the observable effects of the parameters in numerical examples and with real data. Finally, section \ref{conclusion} gives some final thoughts.

\section{The LASSO operator} \label{lassosec}

Consider a standard linear model of the form
$$
Y=X\beta+\epsilon
$$
where $Y$ is the $1\times n$ data vector, $\beta$ a $1\times p$ vector of parameters, $X^T$ is the design matrix and the errors are $\epsilon\sim\mathcal{N}(0,\sigma^2 I)$. In a situation in which we suspect that many of the coefficients in $\beta$ are 0 the problem of interest is estimating which coefficients are nonzero along with their value. When the dimension of $\beta$ is high in relation to the sample size, classical inference does not work, so this problem becomes a problem of variable selection. For this purpose, one common technique is to use the Least Absolute Shrinkage and Selection Operator (LASSO), see \cite{tibshirani}. In classical statistics LASSO is seen as a likelihood penalization, and in Bayesian statistics it is treated as a Laplace prior \citep{bayeslasso}. In the Bayesian setting, the MAP corresponds with the classical estimator, namely
$$
\hat{\beta}=argmax_\beta \left[\mathcal{L}(\beta,X,Y)-\lambda\sum | \beta_i | \right]
$$
where $\mathcal{L}$ is the Gaussian log-likelihood function. The expression $\lambda\sum|\beta_i|$ is the LASSO operator and it depends on the value of a parameter $\lambda$.

A popular alternative parametrization for LASSO is to write the operator as $k/\sigma^2\sum_{i=1}^p|\beta_i|$, which makes the LASSO estimator equal to
\begin{align*}
    \hat{\beta}&=argmax_\beta \left[-\frac{1}{2 \sigma^2} || Y - X^T\beta ||_2^2 - \frac{k}{\sigma^2} \sum_{i=1}^p | \beta_i | \right]\\
    &=argmax_\beta \left[-\frac{1}{2} || Y - X^T\beta ||_2^2 - k \sum_{i=1}^p | \beta_i | \right].
\end{align*}
Thus, $\hat{\beta}$ no longer depends on $\sigma$.

\begin{figure}
\begin{tabular}{c c c}
\includegraphics [scale=0.9]{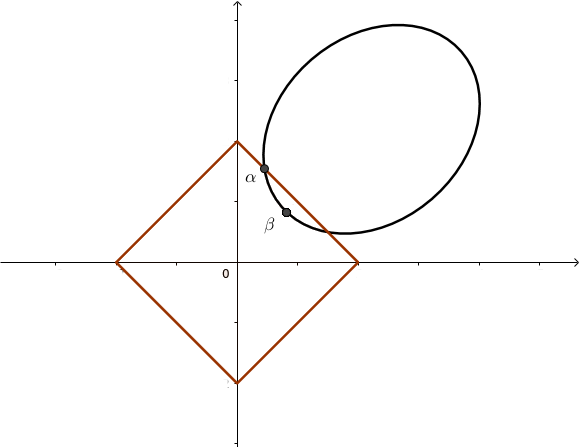} &
\includegraphics [scale=0.9]{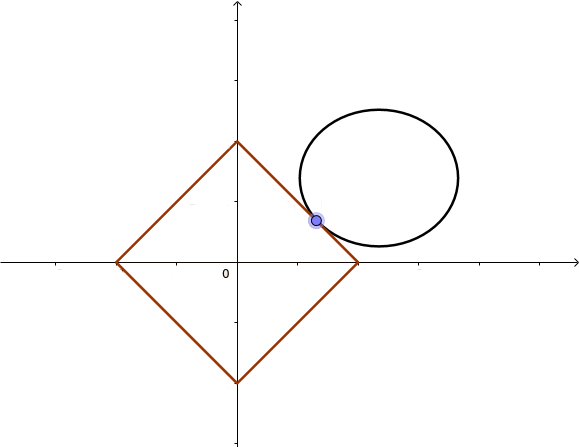} &
\includegraphics [scale=0.9]{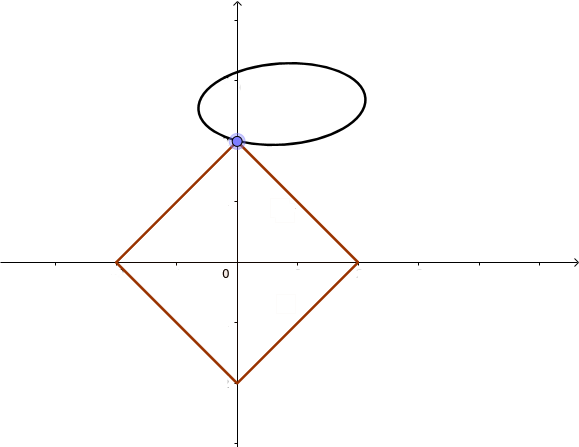} \\
(A) & (B) & (C)
\end{tabular}
\caption{The three forms of intersections of level curves of the likelihood (elipses) and the LASSO operator (squares). (A) cannot happen at the MAP estimator, (B) corresponds to the likelihood curve being tangent to the slope of the prior and (C) corresponds to the likelihood curve intersecting the prior at an extreme, in this case the parameter in the $x$ axis is sent to exactly zero.}
\label{mapcurves}
\end{figure}

The reason why LASSO produces parameter estimations that are \textit{exactly zero} can perhaps best be understood by examining its level curves. In the case where $\beta$ is bivariate there are three possibilities for the geometry of the level curves at the estimator (see figure \ref{mapcurves}). The level curves of the likelihood and the operator may cross (type A), may be tangent (type B), or they might meet at a point at which the curves of the operator are non-differentiable (type C). We note that type A cannot be the estimator by a simple argument: The point marked $\alpha$ cannot be the estimator since at the point marked $\beta$ the value of the likelihood is the same, but the value of the operator is greater. Hence, the estimator must be either type B or type C. It is a $\hat{\beta}$ of type C that interests us given that these situations make the MAP estimator of one parameter exactly equal to zero.

\subsection {How LASSO promotes variable selection}

When considering whether $\hat{\beta}$ is of type B or C, we find that it depends on the value of $\lambda$, and this dependence has a notable property.

\begin{lemma} \label{bivlasso} For fixed X, with probability 1, random data Y will allow the lasso estimator to fulfill the following criterion: There exists $\nu$ such that if $\lambda>\nu$ then $\hat{\beta}$ is of type C.
\begin{proof}
Note that the LASSO operator may be viewed as the Lagrangian for the restricted maximization of the likelihood subject to $\sum|\beta_i|<t$ for some $t$. The larger the value of $\lambda$, the smaller the value of $t$, and when $\lambda\rightarrow\infty$ then $t \rightarrow 0$.

For the bivariate case, consider the slope of the level curve of the likelihood function at the origin. With probability 1, this slope will be neither 1 nor -1. 

Note that the level curves of the likelihood function are concentric. Hence, there is an open ball around the origin where the level curve does not have a slope of 1 or -1 either. In this area, it is impossible for $\hat{\beta}$ to be of type B. Therefore, for large enough $\lambda$, $\hat{\beta}$ must be of type C.

Now for the general case, note that all of the bivariate marginals behave as the bivariate case just explained.
\end{proof}
\end{lemma}

\begin{theorem} For a fixed X and a fixed random Y, with probability 1 there exists $\nu$ such that if $\lambda>\nu$ then $\hat{\beta}_i=0$ for all except one value of $i$.
\begin{proof} 
$\nu$ is the maximum threshold for each pairwise comparison of $\beta_i$ vs $\beta_j$.
\end{proof}
\end{theorem}

In other words, there is almost certainly a threshold for which any $\lambda$ above this threshold will make all estimators zero except for one.

In general, there is no clear way to choose $\lambda$ so as to select variables in any controlled way. In other words, we know that when $\lambda$ grows, our selection becomes tighter and tighter, discarding more and more variables, but there is no interpretable measure of how \textit{much} tighter. In other words, the choice of $\lambda$ can run the gamut from allowing all coefficients to be nonzero to allowing only one of them, and no good way to control its degree of selectivity.

In practice, the most common method for selecting $\lambda$ is to use data-driven techniques such as cross-validation \citep{lassocv, nonlinlasso}.

In passing, we note the following important point: A known issue with LASSO is that estimations depend on the scale of the variables, so it is common practice to center the covariates and standardize them so that $\sum_ix_i^2=1$ \citep{nonlinlasso}, although recently there have been alternative suggestions on how to rescale the variables \citep{rescale}. Regardless of the specific method, something must be done unless the scale of the covariates is carefully chosen. This point is critical not only in LASSO, but in other selection operators as well. For this paper, we will assume that prior to any regularization, covariates have been centered and standardized in the way described above. This will become important when performing calculations related to our proposal later on, but it is equally critical in LASSO, so we mention it now. 

\section {An alternative proposal} \label{introfatso}

\begin{figure}
    \centering
\includegraphics[scale=0.9]{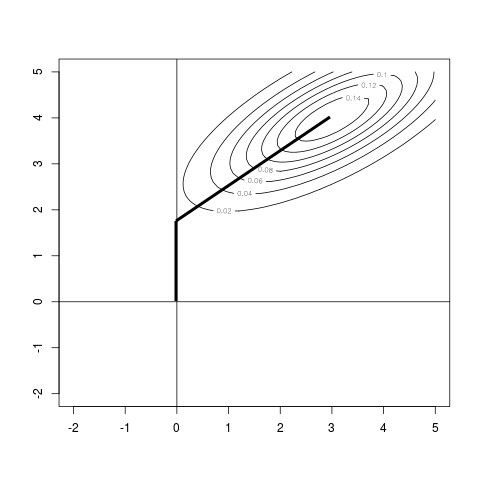}
\caption{The possible locations for the LASSO estimator, as determined by the level curves. Which specific location corresponds to the LASSO estimator depends on $\lambda$. The dark line runs from the MLE along the points where the level curves are at a 45 degree angle, until it reaches an axis.}
\label{trajectory}
\end{figure}

We note, as seen in the proof of lemma \ref{bivlasso}, that the behavior of the level curves of the likelihood at zero is directly related to what variable selection choice will be made by the LASSO estimator. Essentially, the LASSO estimator will be either at a point where the level curves of the likelihood are at a 45 degree angle, or it will make a selection. The only time that it will select both variables regardless of $\lambda$ is if the likelihood level curves are at a 45 degree angle exactly at 0 (for Gaussian data, the probability of this occurring is zero). In figure \ref{trajectory} we see a graphical representation of exactly where the LASSO estimator may be located (depending on choice of $\lambda$).

In order to address the issue of selectivity, we propose to alter the LASSO level curves. The idea is to propose a new set of level curves directly, and to build a selection operator from this proposal. The objective is to adjust the slopes of the level curves such that they span a continuous range. If the slope of the likelihood level curves at zero is in this range, then one variable will not dominate the other.

If the slope of the level curve at zero is in this range then $\hat{\beta}$ will be of type B regardless of the degree of shrinkage.

\begin{figure}
    \centering
    \begin{tabular}{cc} 
        \includegraphics[scale=1.5]{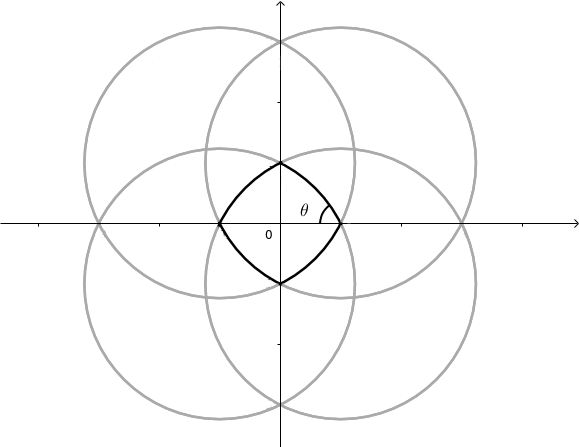}&
        \includegraphics[scale=0.105]{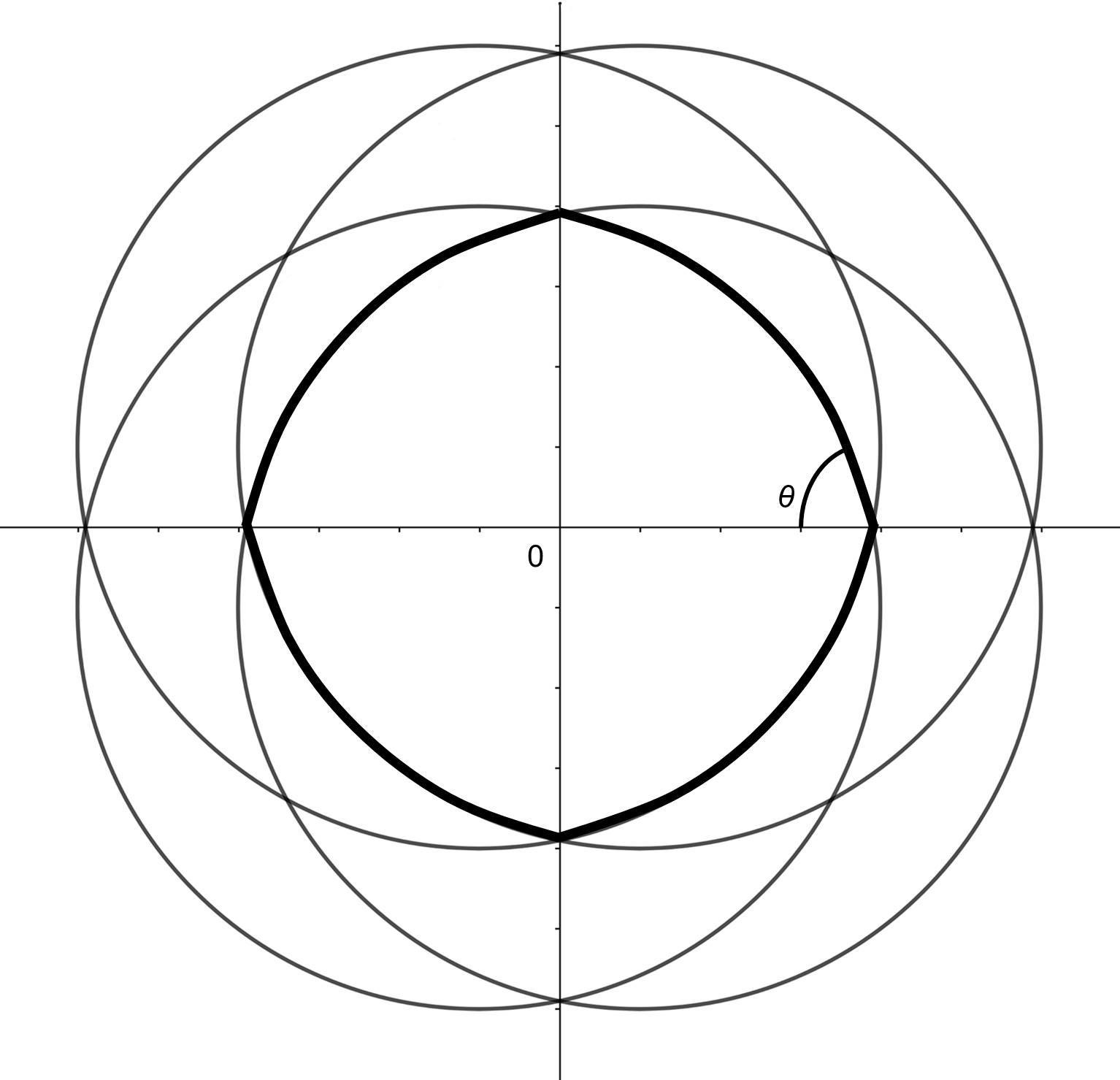}
    \end{tabular}
\caption{The proposed operator's level curves are the boundary of the intersection of disks. If the angle of the likelihood level curves at 0 falls between $\theta-\frac{\pi}{2}$ and $\theta$ then one variable will not dominate the other regardless of the level of shrinkage ($\lambda$ in LASSO). We will resort to the angle $\theta$ introduced in this figure, and shown again in figure~\ref{triangulos}, in many parts of the paper. The value of the level curve corresponds to the level of shrinkage, but a new parameter $\rho$ is introduced, to change the geometry and the angle $\theta$, which controls the position and size of the circles. The two images show the geometry with a different $\rho$ and $\theta$.}
\label{circulitos}
\end{figure}

The geometry of the proposed level curves is the perimeter of the intersections of disks, as illustrated in figure \ref{circulitos} (or in general the boundary of the intersection of $d$-balls in dimension $d$). If the angle $\theta$ in the figure is the same for all level curves, then if the angle of the likelihood level curves at the origin is between $\theta-\frac{\pi}{2}$ and $\theta$, then both variables will be selected regardless of the degree of shrinkage (parameter $\lambda$ for LASSO). This construction will introduce a second parameter $\rho$, which determines $\theta$, and which will be used in addition to a shrinkage parameter.

For the construction to make sense, the angles of intersection of the level curves with the axes must not depend on the degree of shrinkage. Consequently, the center of the corresponding circle will vary depending on which level curve we are on. We proceed to explore the necessary calculations for the construction of an operator from this idea.

\begin{figure}
    \centering
\includegraphics[scale=1.5]{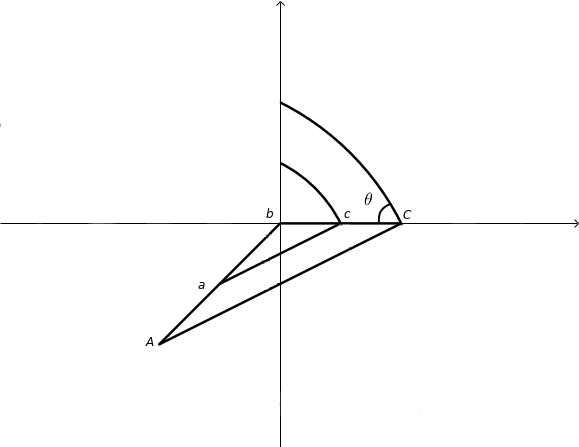}
\caption{The geometry required for calculating the value of the operator. $A$ and $a$ are the centers of the circles , the arcs of which intersect the horizontal axis at $C$ and $c$, respectively. Note that triangles $abc$ and $AbC$ are similar. This figure is a reference for several calculations throughout the paper.}
\label{triangulos}
\end{figure}

Figure \ref{triangulos} shows the essential geometry used to calculate the location of the center of each curve. We note that triangles $abc$ and $AbC$ share intersection $b$ and we also note that the angle at $c$ is the same as the angle at $C$ so these triangles are similar. We can therefore characterize the angle $c$ by $\rho=||ac||/||ab||=||AC||/||Ab||$. We can now write $a=a_\beta=\alpha \boldsymbol{1}$ where $\boldsymbol{1}$ is a $1\times p$ vector of ones, and $\alpha$ is the distance from $a$ to the origin along any given axis. $\rho$ is the additional parameter in our operator, which will be directly related to the desired level of selectivity.

With this notation, and using $d$ for the dimension of $\beta$ it is now possible to write out the calculation
$$
||ac||=\rho||ab||
$$
$$
\sqrt{\sum_i(|\beta_i|+\alpha)^2}=\rho\sqrt{p\alpha^2}
$$
$$
\alpha^2 \left( p \left[ 1-\rho^2 \right] \right)+\alpha \left(2\sum|\beta_i| \right)+\sum|\beta_i|^2=0.
$$

In the range of interest, $\rho>1$ and $\alpha>0$ so we solve this equation to find a closed form expression for $\alpha$
$$
\alpha=\frac{-2\sum|\beta_i|-\sqrt{\left(2\sum|\beta_i|\right)^2-4\left(\sum|\beta_i|^2\right)\left(p\left[1-\rho^2\right]\right)}}{2\left(p\left[1-\rho^2\right]\right)}.
$$

We must remember that in this section we have written $a$ and $\alpha$ out of notational convenience, but that they depend on $\beta$ and on $\rho$, so it really is $\alpha_{\beta,\rho}$ and $a_{\beta,\rho}$.

Now that we have computed the geometry of the problem, the remaining issue is to use this geometry to construct an operator (in this case one that will also match a prior distribution). Any probability distribution for which the level curves of the density function are concentric circles (or higher dimensional equivalent) centered at the origin may be used as a basis for the construction of the operator. If the density function is $f(\beta)$ then we can construct a distribution with density function $g(\beta)\propto f(|\beta|+a_{\beta,\rho})$. This does not have a scale parameter unless $f$ does, but most useful distributions do have one. We will refer to this family of priors as FATSOs or Flexible Axis-Thickened Selection Operators, and the basic form of FATSO will be based on a Gaussian distribution.

    The FATSO will always be a probability distribution so long as $f$ is also a probability distribution, and will have finite moments whenever $f$ does since
$$
\int h(\beta)g(\beta)d\beta\propto\int h(\beta)f(\beta+\alpha_{\beta,\rho})d\beta\leq\int h(\beta)f(\beta)d\beta.
$$

The full formula for the Gaussian FATSO will have log density

$$
log [g(\beta)]=K_{\rho,\lambda}+\lambda\sum(|\beta_i|+\alpha_{\beta,\rho})^2
$$
for some normalizing constant $K$, which does not have to be computed since it does not depend on the $\beta_i$s.

This distribution also has the following useful property:

\begin{lemma} The negative log density of the Gaussian FATSO is concave.

\begin{proof}

Note that $-log(\phi(x))$ (where $\phi$ is the univariate Gaussian density) is an increasing function for positive $x$. Note also that $-log[g(\beta)]=-log(\phi(\beta+\alpha_{\beta,\rho}))$. We also observe that $\alpha_{\beta,\rho}$ is convex when seen as a function of $\beta$, so the result follows.
\end{proof}

\end{lemma}

A trivial corollary is that, since the likelihood function for linear regression is also log-convex, then the posterior is log-convex and the calculation of the MAP is a convex optimization problem. Unfortunately most convex optimization algorithms require the use of gradients, and FATSO is not differentiable at any point where some $\beta_i=0$, so the gradient does not exist at the expected optimum. That said, the convexity of  the target function guarantees a unique maximum, and other desirable properties for optimization. Almost any optimization technique which does not depend on differentiability at the optimum will calculate the FATSO estimator effectively.

\section{Interpreting FATSO and selecting parameters} \label{rho} 

The design of FATSO is based around the idea of reducing the collection of level curves for which parameter estimates are zero in a controlled way. Namely, the issue is the slopes of the level curves of the likelihood function at zero. By adding the parameter $\rho$, we have allowed an interval of these slopes to produce nonzero parameter estimates, rather than a single slope. This seems promising, but in order to be of real use, we need a proper way to interpret this slope and assign $\rho$ (and $\lambda$ in most cases) to fit our problem.

As we have previously observed, in the bivariate case, if the angle of the level curves is between $\theta$ and $\frac{\pi}{2}-\theta$ then both variables will be selected.  For interpretative purposes, let $m$ be the slope $m=tan(\theta)$; $\theta$ as in figure~\ref{circulitos}. Following the geometry from figure \ref{triangulos} we can observe that $b$ is a known angle $\left(\frac{3\pi}{4}\right)$.  Using $sin(b)=\frac{1}{\sqrt{2}}$ allows us to calculate
$\rho=\frac{|AC|}{|Ab|}=\frac{sin(b)}{sin(c)}=\frac{1}{\sqrt{2}}$ and $\theta=cosec^{-1}(\sqrt{2}\rho)$, so we have
$$
\rho=\sqrt{\frac{1+m^2}{2}} .
$$

We now have an easy conversion between $\theta$, $m$ and $\rho$, but on its own this brings us no closer to interpreting $m$ nor to being able to set $m$ (ie. $\rho$) in the FATSO operator.

The key to this crucial step is to calculate the slope of the likelihood level curve at zero. We note that the level curves are perpendicular to the gradient, so it is possible to study this slope by considering the gradient of the likelihood function at zero.

We observe that for the standard linear regression problem, the likelihood function is integrable, and a flat prior can be used to obtain a Gaussian posterior \citep{bayeslinmod}. We will not actually use a flat prior nor treat the result as a posterior, but for mathematical convenience, we can think of the likelihood function as if it were a Gaussian density $\pi(\beta)$ with mean $\mu$ at the MLE and covariance matrix $\Sigma=\left(X^TX\right)^{-1}\sigma^2=\left[\begin{array} {cc}\varsigma_i^2&\varsigma_{ij}\\ \varsigma_{ij}&\varsigma_j^2\\ \end{array}\right]$.

The gradient of $\pi(\beta)$ of a Gaussian density is \citep{matrix}
$$
\frac{d\pi(\beta)}{d\beta}=-\pi(\beta)\Sigma^{-1}(\beta-\mu) .
$$
When we reduce it to the bivariate case, the slope of the gradient at zero is
$$
\frac{\mu_j\varsigma^2_i-\mu_i\varsigma_{ij}}{\mu_i\varsigma^2_j-\mu_j\varsigma_{ij}}.
$$
As previously explained, the covariates have been standardized so $\sum_ix_i^2=1$ and hence it is easy to observe that $\varsigma_i=\varsigma_j$, so we can write this quantity with $\varsigma$, obtaining
$$
\frac{\mu_j\varsigma^2-\mu_i\varsigma_{ij}}{\mu_i\varsigma^2-\mu_j\varsigma_{ij}} .
$$

In the independent case, where $\varsigma_{ij}=0$ (which can only happen if there is no intercept: If both the covariates and the response variable are centered then the intercept is always 0 anyway) this result is simply the ratio of the signals of the two parameters (note that with standardized covariates these are the pure effects on $Y$, free from the units of measurement; this can be seen easily since $\varsigma_i=\varsigma_j$ so $\frac{\mu_j}{\mu_i}=\frac{\mu_j\varsigma_i}{\mu_i\varsigma_j}$ or the quotient of signal to noise ratios, which are unitless). This corresponds well with an intuitive notion of the relative importance, or difference from zero, of one parameter to the other. In other words, this gives us an interpretation for the slope of the likelihood level curve at zero.

This intuitive notion is quite reasonable in the case of $\beta_i$ and $\beta_j$ are independent, but when they are correlated then it is lacking. If $\beta_i$ and $\beta_j$ tend towards zero together, for instance, then we would hope our notion of relative difference from zero would reflect that.

One way to attempt to correct this is to consider instead the conditional distribution of one variable given the other \citep{gausscond},
$$
\beta_i|\beta_j\sim\mathcal{N}(\mu_i+\frac{\varsigma_{ij}}{\varsigma_j^2}(\beta_j-\mu_j),(1-\frac{\varsigma_{ij}^2}{\varsigma_i^2\sigma_j^2})\varsigma_i^2),
$$
and calculate the conditional equivalent which we will call $r_{ij}$

$$
r_{ij} = \left|\frac {\mathbb{E}(\beta_i|\beta_j=0)} {\mathbb{E}(\beta_j|\beta_i=0)}\right|.
$$

This would give a more accurate representation of the relative difference from zero of the two variables, since it is the quotient of the means in the particular case of interest in which the other variable is zero (Also, $var(\beta_i|\beta_j=0)=var(\beta_j|\beta_i=0)$ so this is still unitless).

\begin{figure}
    \centering
    \begin{tabular}{cc}
        \includegraphics[scale=0.4]{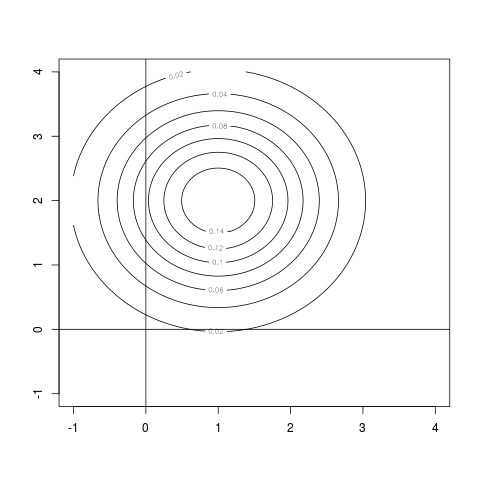} &
        \includegraphics[scale=0.4]{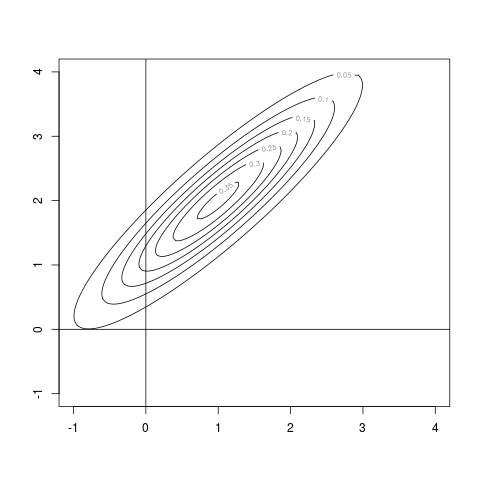}
        \\
        (A) & (B)
    \end{tabular}
\caption{\label{gausscurves} Two Gaussian bivariate densities with the same marginals and different degrees of correlation. In (A) the variables are independent and  (with the vertical variable as $\beta_i$ and the horizontal variable as $\beta_j$) we have $r_{ij}=2$. In the second case, the two variables are strongly correlated. When one variable tends towards zero, the other is also very small. In this case, intuitively, the variables are closer and there is less of a reason to prefer one over the other. This intuition is reflected by $r_{ij}=1.375$.}
\end{figure}

Figure \ref{gausscurves} gives some intuition to show how the conditional distribution is a better choice than the marginal distribution. Both of the Gaussian distributions shown have the same marginal density, but in one case they are independent and in the other they are highly correlated. The difference between the relative importance of the two variables is visually apparent: If one of the variables is set to be zero, the other should be small as well.

When we calculate $r_{ij}$, the result is exactly $\left|\frac{\mu_i\varsigma^2-\mu_j\varsigma_{ij}} {\mu_j\varsigma^2-\mu_i\varsigma_{ij}}\right|$, which is precisely the slope of the gradient of the likelihood at zero.

In other words, regardless of correlation, the slope of the level curves of the likelihood at zero matches the conditional signal ratio, $r_{ij}$, which is a good intuitive measure of relative importance between variables in a regression problem.

The user therefore assigns $m$ as the circumstances require so that, regardless of $\lambda$, both $\beta_i$ and $\beta_j$ are selected if $r_{i,j}$ is between $m$ and $\frac{1}{m}$. Our previous calculation allows us to set $\rho$ when $m$ is known, although it is also possible to simply use an alternate parametrization, working with $m$ directly instead of $\rho$. This parametrization is easier to interpret and will be used from here on out.

This is nicely interpretable in two dimensions. In higher dimensions the structure is analogous and the mathematics are identical (Simply do the calculation with the marginal distribution of the two intended variables). The interpretation of the slope is slightly less intuitive, since the direction is determined by a vector rather than by a single number. The relationship of the corresponding components of the gradient, however, still matches $r_{i,j}$.

We have a way to interpret $m$. For full Bayesian inference, one would simply select $m$ but it may also be reasonable to choose another path and simply try out values of $m$. Since the computational cost is low (unless the number of parameters is truly huge), a fair amount of information about the behavior and relative importance of parameters can be gleaned in fairly little time. In table \ref{results2}, from section \ref{secresults}, we can see an example of what such an exploration might look like.

One final practical note on the subject of the selection of $m$ is based on the fact that it is independent of units. Since it means the same at all scales, one can think that reasonable values for $m$ should be between 1.1 and 15. If $m$ is less than 1.1 then there is very little difference between the geometries of FATSO and LASSO, whereas if $m$ is greater than 15 one is hardly performing any variable selection at all.

\subsection{$\lambda$ and prior conditional variance in Gaussian FATSO}

We now have a handle on $m$ but Gaussian FATSO has a second parameter $\lambda$. If $\lambda\rightarrow 0$ then we end up with a flat prior which may be suitable to variable estimation without any selection. On the other hand if $\lambda\rightarrow \infty$ then the prior will be concentrated around zero. This yields higher selectivity, but also shrinks the value of all estimations.

One way to think about selecting $\lambda$ is to think of the FATSO less as an operator and more as a prior distribution. We can then study the properties of FATSO as a probability distribution, in which case $\lambda$ may be interpreted as related to the variance of this distribution. Following the geometry from figure \ref{triangulos} we have the next lemma.

\begin{lemma}For a Gaussian FATSO, the prior distribution of $(\beta_i|\beta_j=0\forall j\neq i)$ is a Gaussian random variable with mean zero and (prior)variance

$$
\lambda^{-1}\sqrt{2}sin\left(\theta-\frac{\pi}{4}\right)
$$
    where $\theta$ is the angle as shown in figures \ref{circulitos} and \ref{triangulos}.
\begin{proof}
We note from figure \ref{triangulos} and Tales's theorem, that the ratio of AC to BC is the same regardless of how far C is along the horizontal axis. Then we have the relationship
$$
\lambda||C-A||^2=\lambda k||C-B||^2
$$
where the left hand side differs by a constant from the value of the FATSO prior log-density calculated at the point C and the right hand side differs by a constant from a Gaussian log-density calculated at the same location (B is the origin).

Some trigonometry will then yield the value of $k=\sqrt{2}sin(\theta-\frac{\pi}{4})$, which proves the claim.
\end{proof}
\end{lemma}

When $\theta\rightarrow\frac{\pi}{4}$ then the variance goes to zero, and the geometry of FATSO approaches the geometry of LASSO.

Of note, if $\theta$ is close to $\frac{\pi}{4}$ then $sin\left(\theta-\frac{\pi}{4}\right)$ can become very small, and as a result $m$ will have an effect on shrinkage of estimators unless $\lambda$ is adjusted to compensate. This is not a very significant issue unless $m<1.1$

If FATSO will be used for Bayesian analysis, this shows the effect of $\lambda$ on the FATSO prior. The conditional variance of 
one parameter given all others are zero is a reasonable way to establish prior variability. $\lambda$ should be selected accordingly.

Departing from a full Bayesian prior statement, one reasonable way to select $\lambda$ is to use data-driven techniques such as cross-validation, but these may come at a significant computational cost, or the sample size may be too small for cross-validation to be a reasonable choice.

If we want to set $\lambda$ using heuristics, we will turn to the observed data $Y$ for some guidance. Note that if only $\beta_i$ is active and all others are equal to zero, then if we write $X_i$ as the $i$th column of X we have
$$
Y=X_i\beta_i+\epsilon.
$$

Now, using the Bayesian interpretation (even if we are not going to perform Bayesian inference), we can think of $\beta_i$ as a random variable (a priori independent of $\epsilon$) and write

$$
var(Y)-\sigma^2=var(\beta_i)X_i^TX_i,
$$
and here we use the fact that X was standardized so that $\sum_jX^2_{j,i}=1$.

$var(Y)$ is not known, but it can be estimated with the sample variance $var(Y)\approx\sum \frac{(Y_i-\bar{Y})^2}{n}$. Hence, if $var(\epsilon)$ is known we can calculate one choice for $\lambda$ as follows
$$
\lambda=\sqrt 2 sin\left(\theta-\frac{\pi}{4}\right)[var(Y)-\sigma^2]^{-1}
$$

However, it is worth observing that in practice, the value of $\lambda$ has a relatively small effect on point estimates, as will be seen empirically in the results section of this paper. $\lambda$ acts more as an on/off switch than a dial, and hence it is not too important to worry about its exact value. One has only to find something in the (usually very large) reasonable range. The above method for selecting $\lambda$ is not meant to be taken as a precise value, but merely to give a notion of where the reasonable range might be.

A second option, as is done in LASSO, mentioned in section \ref{lassosec} is to parameterize not with $\lambda$ but with $k/\sigma^2$, yielding estimates which no longer depend on $\sigma$. Of course, this comes at the cost of being able to use knowledge of $\sigma$ in order to select the parameter, as was done above with $\lambda$. Even if we prefer to use $\lambda$, however, this shows us that we can scale $\lambda$ with the inverse of the standard deviation of the noise to achieve similar results.

\section{Comparison to other LASSO extensions} \label{comparisons}

FATSO is not the first attempt to extend the ideas of LASSO in a new direction. There are several other regularizations which have been attempted and which yield different benefits. We make no claims that FATSO is necessarily any better than any of these, but only that the issues it aims to address are different.

\subsection{Ridge and Bridge regression}

Ridge regression, also known as \textit{Tikhonov regularization} in inverse problems \citep{inverse} and is an older idea than LASSO. It is also closely related to the use of Gaussian priors in Bayesian regression. It essentially aims to estimate the regression coefficients with
$$
\hat{\beta}=argmax_\beta\left[\mathcal{L}(\beta,X,Y)-\lambda\sum\beta_i^2\right],
$$
where $\lambda\sum\beta_i^2$ is the Ridge operator. One idea which places LASSO at one end and Ridge regression at the other is called Bridge regression, which changes the operator to $\lambda\sum|\beta_i|^\alpha$ for $\alpha\in(1,2)$. Of note, however, for any value of $\alpha>1$ the slope of the level curves at 0 is exactly zero. Hence, Ridge and Bridge are not selection operators in the sense that the resulting estimators are not zero \citep{ridge, bridge}.

\subsection{Group LASSO}

One common extension to LASSO is the group LASSO, which separates the columns of X into groups and which promotes the selection of groups of variables together. While this does extend the ability of LASSO to handle more complex situations, it also requires some degree of understanding of the relationships between covariates, which is not the goal of FATSO. In another sense, however, group LASSO is more closely related to FATSO than the other LASSO extensions since it aims to incorporate information about parameter grouping that is not immediately visible in the data but which is understood by the user \citep{grouplasso}.

\subsection{Scale mixtures of Normals}

In recent Bayesian literature, there has been an explosion of selection operators proposed with the theme of corresponding to priors which are scale mixtures of normals \citep{horseshoe, otherscalemix, mixgpriors}. A scale mixture of normals is a random variable $X$ which can be represented as $X=Y\sigma$ where $Y$ is a random variable with a standard normal distribution and $\sigma$ is some other (continuous or discrete) random variable \citep{scalemix}. LASSO itself is closely related to this family, since it corresponds to a Laplace prior and a single variate Laplace prior is a scale mixture of normals with $\sigma$ a Gamma distributed random variable. There are various motivations for the proposed operators, but they generally are focussed on some form of asymptotic convergence either of the entire posterior distribution or of some point estimate derived from it. We are unaware of any which ease the interpretation of tuning parameters.

\subsection{Elastic net}

The idea with the most similar behavior to FATSO is the elastic net. The elastic net uses as a regularization operator $\lambda_1\sum|\beta_i|+\lambda_2\sum\beta_i^2$ (and then applies a correction to the estimator), essentially working as a linear combination of the Ridge and LASSO operators. The first thing to note about the elastic net operator is that the level curves are not concentric, and the slope of the curves' intersection with the axes depends on the curve. For distant curves, the geometry of Ridge is dominant, whereas with curves closer to the origin the geometry is closer to that of LASSO.

While elastic net does not maintain the concentric level curves of FATSO, it does allow for variable selection with less stringent selectivity than LASSO, so it behaves in a similar way. In elastic net, however, the degree of selectivity is moderated very obscurely by the interplay of $\lambda_1$ and $\lambda_2$. The common recommendation is to select both parameters by data-driven techniques, such as cross-validation. This is a valid approach, but does not allow users to make informed decisions about the desired degree of selectivity based on their own expertise. Given that the $p>n$ scenario is one where data is known to have very little information, the goal of allowing human knowledge to participate is very sensible.

While FATSO is in no way intended to replace the elastic net, it is worth noting that the two main issues with LASSO which the elastic net aims to solve are both addressed by FATSO as well. The first of these issues is that in $p>n$ situations, LASSO cannot select more than $n$ variables, and in the following section we will see an example where FATSO selects more than $n$ covariates. The second issue is that when several covariates are highly correlated LASSO tends to select only one of them. It is proven in the original elastic net paper \citep{elasticnet} that any strictly convex regularization will solve this issue and FATSO is strictly convex.

FATSO does not aim to compete with the elastic net in terms of computational tractability or in terms of asymptotic error reduction, so while the behavior of the two operators is somewhat similar, their ultimate objectives are different.

\section{Results} \label{resultsec}

\subsection{Simulated Data}

15 observations of a 20 dimensional regression problem were simulated. The true values of the regressors $\beta$ were zeros except for three variables. These variables were $\beta_1=0.9$, $\beta_{2}=0.5$, and $\beta_{3}=0.7$ (noise standard error was 0.1). Using these same data, FATSO estimates were calculated using different values of $\rho$ and $\lambda$. Table \ref{results} shows maximum a posteriori estimates of these data based on various values of $\rho$ and $\lambda$ using Gaussian FATSO.

\begin{table}
\caption{\label{results}Results of estimations using the FATSO operator using various values for $\rho$ and $\lambda$. We see that $\rho$ affects selectivity smoothly, while adjusting $\lambda$ does not allow for very flexible tuning.}
\begin{tabular}{|c|c|c|c|c|c|c|}
\hline
$m$ & $\rho$ & $\lambda$ & $\hat\beta_1$ & $\hat\beta_{2}$ & $\hat\beta_{3}$ &Other $\beta$s\\
\hline
\hline
500 & 353.6 & 0.1 & 0.697 & 0.329 & 0.543 & Many are of similar\\&&&&&& order of magnitude\\
\hline
    30&21.21&30&0.709&0.336&0.548&$|\beta_6|,|\beta_{7}|,|\beta_{12}|,|\beta_{16}|, |\beta_{20}|$\\&&&&&&also active\\
\hline
3 &2.236 &30 &0.814 &0.397 & 0.602 & $\beta_7$ and $\beta_{17}$ are active\\
\hline
2&1.581&30&0.871&0.464&0.671&Others inactive\\
\hline
2&1.581&0.01 &0.857&0.442&0.65&$\beta_{17}$ active\\
\hline
2&1.581&1&0.868&0.459&0.69&Others inactive\\
\hline
2&1.581&100&0.868&0.46&0.658&Others inactive\\
\hline
2&1.581&1000&0.71&0.249&0.437&Others inactive\\
\hline
\end{tabular}
\end{table}

With a high value for $m$ and a low value for $\lambda$, the FATSO prior is nearly flat. In these situations the estimator is nearly the MLE, and since the dimension is greater than the sample size, the MLE does not give any real information about the parameters. As FATSO becomes more informative, so do the estimations. Similarly, we note the effect of $\lambda$ and $m$ act independently. We note that the effect of $\lambda$ acts almost as if it had a threshold. For a fixed $m$ of 2, then for any $\lambda\geq1$ the exact value does not seem to have very much effect. $\lambda$ at 1 and at 100 both yield very similar estimations for the parameters, and it is not until $\lambda$ is extremely large (1000) that the effect of shrinkage becomes noticeable. With very small $\lambda$, however, the effect is lost somewhat (the extreme case being $\lambda=0$ where we are left with the MLE again). The same cannot be said for $m$ (or $\rho$), which affects estimations much more fluidly. We see that with a fixed $\lambda$ of 30, the effect of $m$ on selectivity is very clear. As $m$ approaches 1 the selection is more strict, and as $m$ grows then selection is looser. This confirms that the parameter $m$ permits tuning the degree of selectivity in a fluid way that is not possible with a shrinkage parameter alone.

It is very tempting to be seduced by good results with $m=2$ and reasonably high $\lambda$ since the estimations are so close to the truth, but we must remember that these are synthetic data. The estimates with higher values of $m$ are also estimates that could easily produce the same dataset, but in this particular case did not. When the dimension of the parameter space is larger than the sample size, then the data will be in the column space of the design matrix. The choice of one set of estimators over another is not information that is really in the data at all. With lower $m$, FATSO will tend to choose smaller sets of covariates which explain the data, but whether that is desirable or not is really an issue of user to judge.

In order to illustrate this case, data was simulated with 17 nonzero variables rather than 3. It is known that LASSO selects at most as many variables as the sample size, so using LASSO here will select for at most 15 of the 17 active variables. The variables that were zero were $\beta_1$, $\beta_{11}$ and $\beta_{19}$ and inference was performed in the same way. With $\rho=1.2$ and $\lambda=100$ (a highly selective combination with a geometry similar to LASSO) we estimate 6 inactive variables. These are $\beta_4$, $\beta_6$, $\beta_{10}$, $\beta_{11}$, $\beta_{12}$, and $\beta_{19}$. As we see, not only are the variables being selected more strictly, but the variable choice is simply wrong. This is caused by the insistence on a high level of selectivity. With $\rho=3$ and $\lambda=100$ we estimate two inactive $\beta$s, and these are $\beta_{11}$ and $\beta_{19}$. In this case the random simulation turns out to be unusually highly correlated with $\beta_1$ simply by chance, so $\beta_1$ was not selected against. We note that specifying too stringent selection criteria forces the model to shift away from the true values of the parameters, but allowing more nonzero entries yields a very good selection of variables. The difference between this situation and the last one is subtle, and it may often be a good idea to make the selection based on human understanding of the situation rather than on data which is necessarily insufficient.

\subsection{Real Data} \label{secresults}

We use data from \cite{prostate} to evaluate the performance of FATSO and the effect of parameter adjustment. These data are often used for LASSO demonstrations. The data are a 9 column matrix which describe prostate cancer data in 97 patients. The first 8 columns describe characteristics of the tumor and the last column is the response variable: An antigen. The data is available in the R package \textit{lasso2} \citep{lasso2}

Since the data was sorted by the response variable, the rows of the matrix were permuted randomly. There are 97 rows; estimation was done using the first 67 rows and used to estimate the remaining 30. Table \ref{results2} displays the results of inference on this collection of data using different values for the parameters.

\begin{table}
\caption{\label{results2}Results of estimations using the FATSO operator on the prostate cancer data. Once again we see that $\rho$ can be adjusted to tune the degree of selectivity with some reasonable degree of control.}
\begin{tabular}{|c|c|c|c|c|c|c|}
\hline
$m$ & $\lambda$ & active $\beta$s & Prediction MSE &Observations\\
\hline
    100 & 0.0001 & all except $\beta_8$ &0.60516 & Almost exactly \\&&&& simple linear regression \\
\hline
5&0.5& all except $\beta_8$ &0.61363& \\
\hline
    3 &1& $\beta_1, \beta_2, \beta_4, \beta_5, \beta_6, \beta_7$ &0.63189& Removing one variable \\&&&& does not greatly \\&&&&increase the error \\
\hline
2 &1 & $\beta_1, \beta_2, \beta_4, \beta_5, \beta_6, \beta_7$ &0.65002 & \\
\hline
1.2 &1 & $\beta_1,\beta_3, \beta_6$ & 0.78280 & With a higher error\\&&&& we can remove many more\\
\hline
\end{tabular}
\end{table}

The main takeaway from this experiment is the fluid way in which we can select the $\beta_i$s. Since this is not a high dimensional problem, the MLE is a good estimator, but by tuning $m$ we can pick a simplified model which selects more or fewer variables. Removing variables comes at a cost, but we can see exactly how costly this removal is. Using this information it is possible to manually tune our model to whatever balance of parsimony and accuracy we want. Hence, even in this relatively low dimensional scenario, there is something to be gained by having a fluid selection operator.

\section{Conclusions} \label{conclusion}

In situations with high dimensional data, where $p>n$, there are infinitely many parameter combinations which might yield the observed data. In these situations, the data does not clearly favor one choice of parameters over another, so in order to make a selection, some measure of human choice is required. LASSO and other similar regularization operators are means by which a form of preference is given to one kind of solution over another. These systems all have parameters which affect -- in some sense -- how this choice is made. The selection of these parameters by data-driven techniques is appealing, but the information to make the choice is not really in the data. As a result, it becomes desirable to understand the meaning of the parameters and the effect of their choice on the resulting inference. This problem is particularly serious in the Bayesian setting since the operators correspond to prior distributions and it is invalid to assign priors using the data that these priors are chosen to analyze. This last issue is not a vague or theoretical one since both LASSO and Elastic Net have been adapted for Bayesian inference regardless of the difficulty in assigning parameters \citep{bayeslasso, bayesenet}.

While significant effort has been made to improve the data-driven techniques for adjusting parameters, this effort has done little in the sense of improving the interpretability of the parameters for human users who have additional information. In this sense, the elastic net is the system which boasts the lowest mean squared error for theoretical purposes, but it is also gives perhaps the least interpretable combination of parameters.

FATSO is an attempt to offer a means of setting the degree of selectivity by hand. While it is theoretically possible to use data-driven techniques to assign $m$, if data driven techniques are preferred, then one would probably be better served using another regularization operator. On the other hand, in situations where one intends to choose the degree of selectivity using outside knowledge, FATSO is recommended to set the selectivity in a way that is understandable and meaningful.

\bibliographystyle{rss}
\bibliography{fatso}

\end{document}